\documentclass[a4paper,11pt]{article}
\usepackage{graphics,bm}
\DeclareGraphicsExtensions{ps,eps,epsf}
\begin{document}
%------------------------------------------------------------
\newcommand{\matr}[1]{\stackrel{\leftrightarrow}{{\bm{\mathsf{#1}}}}}
\newcommand{\Real}[1]{\Re{\rm e}\left[ #1 \right]}
\newcommand{\vep}{\varepsilon}
\newcommand{\me}{\mathrm{e}}
\newcommand{\mi}{\mathrm{i}}
\newcommand{\dif}{\mathrm{d}}
\newcommand{\uvec}[1]{\hat{\bm #1}}
\renewcommand{\vec}[1]{\bm{#1}}
%-------------------------------------------------------------
\title  {Dynamic mechanical response of polymer networks}
%-------------------------------------------------------------
\author {S.\ F.\ Edwards,
H.\ Takano\thanks{ {\normalsize Permanent address:} Department of
Physics, Faculty of Science and Technology, Keio University,
Yokohama 223, Japan } \ and E.\ M.\ Terentjev \\    \\
{\normalsize Cavendish Laboratory, University of Cambridge} \\
{\normalsize Madingley Road, Cambridge CB3 0HE, U.K.}}
%--------------------------------------------------------------
%\date{\today}
%--------------------------------------------------------------
\maketitle
%--------------------------------------------------------------
\begin{center} {\bf Abstract } \end{center}
%--------------------------------------------------------------
The dynamic-mechanical response of flexible polymer networks is
studied in the framework of tube model, in the limit of small
affine deformations, using the approach based on Rayleighian
dissipation function. The dynamic complex modulus $G^*(\omega)$ is
calculated from the analysis of a network strand relaxation to the
new equilibrium conformation around the distorted primitive path.
Chain equilibration is achieved via a sliding motion of polymer
segments along the tube, eliminating the inhomogeneity of the
polymer density caused by the deformation. The characteristic
relaxation time of this motion $\tau_{\rm e}$ separates the
low-frequency limit of the complex modulus from the high-frequency
one, where the main role is played by chain entanglements,
analogous to the rubber plateau in melts. The dependence of
storage and loss moduli, $G'(\omega)$ and $G''(\omega)$, on
crosslink and entanglement densities gives an interpolation
between polymer melts and crosslinked networks. We discuss the
experimental implications of the rather short relaxation time and
the slow square-root variation of the moduli and the loss factor
$\tan \delta (\omega)$ at higher frequencies.
%---------------------------------------------------------------
\vspace{0.5cm}


62.40.+i Anelasticity, internal friction, stress relaxation,

83.50.Fc Linear viscoelasticity
%%%%%%%%%%%%%%%%%%%%%%%%%%%%%%%%%%%%%%%%%%%%%%%%%%%%%%%%%%
\vspace{1cm}
%------------------------------------------------------
\section {Introduction}
%------------------------------------------------------
Viscoelastic properties of dense polymer melts have been
successfully explained by the reptation model approach, where one
examines the constrained motion of a chain within an effective
tube formed by entanglements with other polymers
\cite{degen,Doibook}. In this model, the effect of entanglements
is to restrict the lateral motion of a chain and is modeled, in
the first approximation, by a dynamic constraint in the form of a
tube surrounding the thermally fluctuating polymer. The dynamics
of the polymer is then described by the reptation motion within
such a tube, essentially a diffusion of segments along the random
primitive path, with an effective number of chain segments
constrained by the tube a new dynamic variable. Accordingly, the
characteristic disentanglement time for a chain is that required
for all its length to diffuse out of the initial tube. On the
other hand, the classical theory of rubber elasticity, see for
example Treloar's treatise \cite{treloar} or \cite{deam}, examines
the response of a crosslinked polymer network to an affine
deformation in the opposite limit of high dilution: Only the
topologically quenched random crosslinks are affected by the
strain, while the rest of the chain is assumed free to explore all
its configurational space. Several penetrating attempts have been
made over the years to improve this simple approach and address
the issue of dense, entangled network \cite{gordon,vilgis}.
Although the disentanglement by reptation out of its tube is
impossible in a permanently crosslinked network, the concept of a
dynamic primitive path, around which the polymer segments are
constrained to fluctuate, has been used to describe the static
elastic properties of entangled networks. The sliplink model
\cite{sliplink}, where the entanglements are treated as discrete
overdamped but nevertheless mobile constraints, is also a variant
of the tube approach.

The main success of reptation models in polymer melts has been the
description of dynamic properties and viscoelastic relaxation
rates. Analogous approaches to crosslinked networks are not well
developed at the moment and, in fact, even the basic dynamics of
stress relaxation in rubbers is not described theoretically. Much
progress has been achieved in studies of secondary effects, such
as the relaxation of free dangling ends attached to the network,
leading to a slow power-law decay of stress \cite{curro,thirion}.
A significant contribution to understanding of the role of
entanglements in rubber elasticity is made by computer simulation
studies, e.g. \cite{everaers}. However, on a basic level, one
first needs to analytically describe the relaxation of an affinely
deformed ideal network. In this paper we investigate the dynamic
mechanical response (the linear complex modulus) of rubbery
networks in the dense limit of high entanglement. One should note
that, unlike in a case of free chains, the entanglements of
networks strands ought to be understood in a more general sense.
In particular, the knots of increasing complexity, described by
corresponding topological invariants, e.g. \cite{kholodenko}, are
not the only way to restrain the lateral motion of network
strands: Even a simple loop, both ends of which are held by
crosslinks, will contribute to the formation of an effective
reptation tube. Therefore, the traditional notion of minimal arc
length required for a polymer to be entangled, cf. \cite{Doibook},
should not be applied to networks -- even relatively short strands
could be highly constrained in a dense system. In such networks,
one expects to find a storage modulus $G'$ even at zero frequency
of imposed deformation (the static modulus $\mu_0$, determined by
the crosslink density in the classical rubber elasticity
\cite{treloar}), while at a finite frequency $\omega$ the response
should be increasingly controlled by chain entanglements in the
similar way as the rubber plateau value of $G'$ in a dense melt.

Let us consider a polymer chain with segments of Kuhn step length
$b$, of which both ends are crosslinked. The entanglement effect
of other network strands is represented by a surrounding tube of
diameter $a$. The central line of such a tube, regarded as a
sequence of taut paths between entanglement points, is given a
name of {\it primitive path} and may be considered as a random
walk of step length $a$, with the end to end distance the same as
that of the polymer chain. In the equilibrium state, the density
of the polymer segments distributes uniformly along the primitive
path. The linear viscoelasticity is then described by the response
of chain segments to the instantaneous deformation. As a starting
point, one may assume that the deformation of the primitive path
is affine, in the sense that both the chain crosslinks and the
entanglement points move in proportion to the macroscopic strain
imposed on the body of rubber. Immediately after such a
deformation, the density of polymer segments along the primitive
path is inhomogeneous, because the deformed primitive path steps
have different lengths depending on their directions and the chain
cannot respond instantaneously to the change of its constraints.
In time, the inhomogeneity of the polymer density relaxes through
the sliding motion of the chain along the new primitive path.

\begin{figure} %[hb]
\centerline{\resizebox{0.84\textwidth}{!} {
\includegraphics{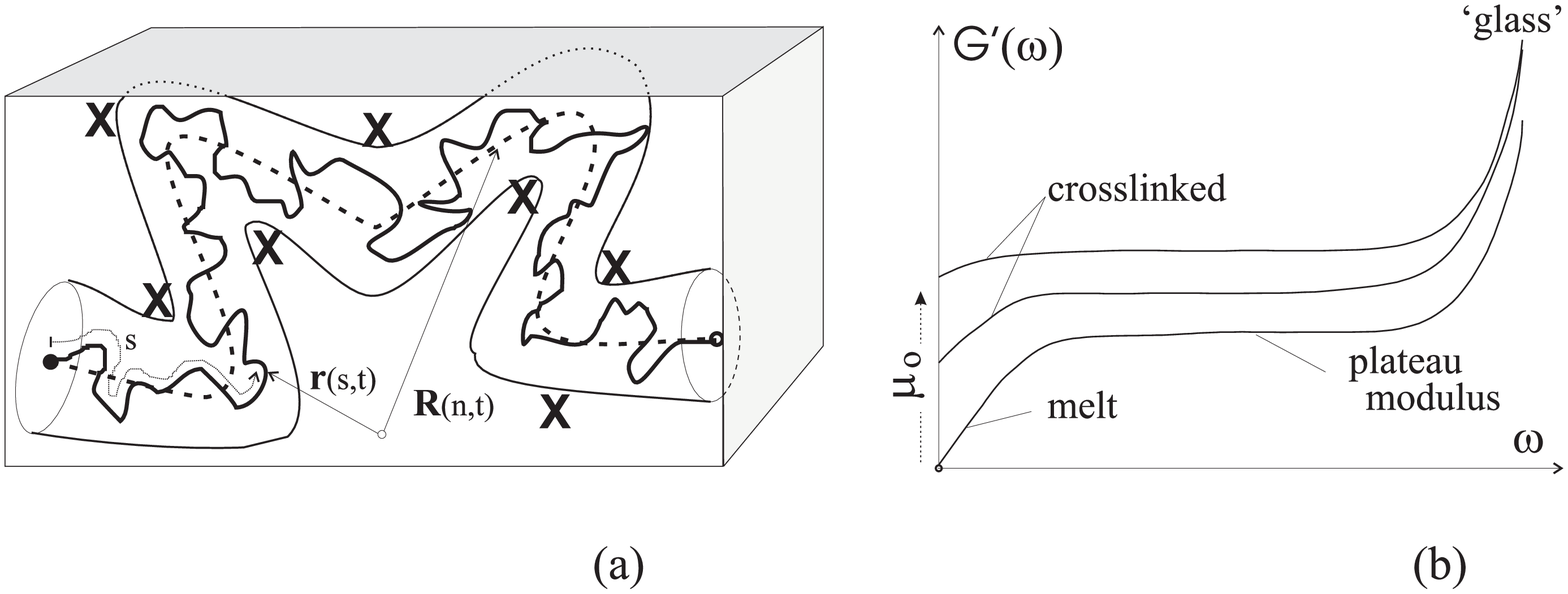} } }
\caption{(a) The scheme of tube model and the relevant variables:
the polymer conformation $\vec{r}(s,t)$ and the tube primitive
path $\vec{R}(n,t)$, where the function $n(s,t)$ specifies the
position along the fixed-length path and $s$ specifies the
position along the chain. In a network, the chain is permanently
crosslinked at both ends of the tube. \ \ (b) The ``expected''
behaviour of the dynamic storage modulus $G'(\omega)$: For an
uncrosslinked melt the static modulus is zero, but reaches a
rubber plateau at a finite frequency due to entanglements. On
increasing concentration of permanent crosslinks the static
modulus $G'|_{\omega=0}=\mu_0$ increases; the plateau modulus has
contributions from both crosslinks and entanglements. At a much
higher frequency all curves should converge to the glass plateau,
which is not the subject of this paper. } \label{tube}
\end{figure}

In a network, the reptation motion of the chain is completely
prohibited by the crosslinking. Therefore, this sliding motion of
chain segments is considered to be more important for the dynamic
properties of crosslinked networks than for corresponding melts,
where the main effect is the reptation diffusion out of the
deformed tube. In this paper, we clarify how the slip motion of
polymer segments along the primitive path affects the dynamic
modulus. The results are summarised in Fig.~\ref{tube}(b). We
concentrate on relatively low frequencies, well below the glass
region. One expects the linear static modulus $\mu_0=G'(\omega
\rightarrow 0)$ to increase with the network crosslink density,
but also with the density of entanglements. In the limit of higher
frequency, $\tau_{\rm e} \omega \gg 1$, with the characteristic
time $\tau_{\rm e}$ of diffusion along the primitive path, the
storage modulus approaches the plateau value corresponding to the
rubber plateau in a melt, where the role of entanglements is
dominant. Our challenge is to develop a universal description that
would also incorporate the limit of zero frequency, $\mu_0$, and
the plateau modulus in a dense melt.

A classical theoretical description of polymer networks is based
on the statistical-mechanical description of Gaussian chains with
topologically quenched constraints of permanent crosslinks, in the
simplest case, and entanglements on a more advanced level
\cite{deam,sliplink}. The computation of a Gibbs partition
function ${\cal Z}=\int \me^{-\beta {\cal H}}$ is insufficient
because of the quenched distribution of crosslinks and the Replica
method has been developed as an extension of classical statistical
mechanics, as one of the ways to evaluate average macroscopic
parameters, such as the free energy $F_{\rm expt} = \langle \log
\, {\cal Z} \rangle$ \cite{deam,Replica}. However, in spite of
many successes of the Replica method, it remains unsuitable for
finding the dynamic properties, such as the complex modulus
$G^*(\omega)$ -- a traditional parameter describing viscoelastic
behaviour of polymers and rubbers \cite{ferry}. One needs a
dynamical, not static equilibrium solution. Here we utilise a
method of solving the dynamical problem of rubber-elastic response
by an extension of Boltzmann approach, based on Langevin and
Fokker-Planck equations. One introduces the Rayleighian ${\cal R}
= L - F + \sum \lambda \, C$ to incorporate the chain equilibrium
properties via its Lagrangian $L$, the kinetic effects via the
dissipative function $F$ (not to be mixed with the free energy!)
and all required constraints $C$ with their corresponding Lagrange
multipliers $\lambda$.

The principles of Rayleigh's dissipative function are described in
some detail in textbooks on analytical dynamics, e.g.
\cite{whittaker,landau}. Introduction of $F=\sum \frac{1}{2}
\gamma_{ik} v_i v_k$, with the velocities $\vec{v} \equiv
\dot{\vec{x}}$ and friction constants $\gamma$, allows the unified
description of Lagrangian systems where the total energy is
dissipated at a rate $\frac{\rm d}{{\rm d}t}E = -2F$, via the
variational equations $$ \frac{\delta}{\delta \vec{x}}\int L \,
\dif t - \frac{\delta}{\delta \vec{v}}\int F \, \dif t =0. $$ For
instance, for a damped one-dimensional motion, ${\cal R}= L - F =
\frac{1}{2}m \, \dot{x}^2 - U(x) - \frac{1}{2}\gamma \, v^2$, and
the variational equation correctly gives $$m \, \ddot{x} - \gamma
\, \dot{x} +
\partial U/\partial x = 0.$$ This method has been proven to remain
valid in problems involving an arbitrary set of Lagrangian
constraints \cite{sfe82,miller} and promises a much greater
mileage in non-equilibrium statistical mechanics of random
networks.

%-----------------------------------------------------------
\section{Tube model and equations of motion}
%-----------------------------------------------------------

In order to describe the sliding motion of a polymer segment along
the primitive path, we define the following variables, see
Fig.~\ref{tube}(a). Let us consider a polymer chain of the arc
length $L$. The arc distance from one end, $s \in [0,L]$, is used
to specify a position along the chain. The conformation of the
polymer at time $t$ is given, as a function of $s$, by the segment
coordinate $\vec{r} (s,t)$. The chain is constrained within a tube
of radius $a$ -- similarly, we take the conformation of the
primitive path, at time $t$, given by the coordinate of tube axis
$\vec{R} (n,t)$, where $n \in [0,Z]$ is used to specify a position
along the primitive path. Now, $Za$ denotes the contour length of
the primitive path and $na$ is the arc distance from one end along
the tube axis (a variable analogous to $s$ for the chain itself).
An essential variable in the problem is the number of chain
segments $s$ contained within one step of the primitive path $n$.
The relation between the position $s$ on the polymer and the
corresponding position $n$ on the primitive path is then given by
a function $n=n(s,t)$. Note that $\vec{r} (s,t)$ and $n(s,t)$ are
dynamical variables of the polymer chain, while the tube
conformation $\vec{R} (n,t)$ is given externally, by the
surrounding constraints.

In order to derive the over-damped equations of motion, we
construct the Rayleighian ${\cal R}$ as a functional of $\vec{r},
\ \dot{\vec{r}}, \ \vec{R}$ and $\dot{\vec{R}}$, omitting the
inertial effect of kinetic energy so that ${\cal R} = - U - F$.
The potential energy term $U$, describing the Wiener random walk
constrained around the path $\vec{R} (n,t)$ of a fixed length, is
given by
\begin{equation}
U= {{3k_{\rm B} T} \over {2b}} \left[ \int_0^L {\left( {{\partial
\vec{r}} \over {\partial s}} \right)}^2 {\rm d}s + q_0^2 \int_0^L
(\vec{r} - \vec{R} )^2 {\rm d}s \right] + \lambda \left[ \int_0^L
{{
\partial n} \over {\partial s}} {\rm d}s -  Z \right] . \label{U}
\end{equation}
Here, $b$ is the step length of the polymer chain and $k_{\rm B}T$
is the Boltzmann temperature energy scale. The potential term
containing $q_0^2$ represents the effect of confining the polymer
segment $\vec{r}(s)$ around the primitive path through a harmonic
potential. The last term in the eq.~(\ref{U}) arises from the
constraint on the primitive path, $ n(L,t)-n(0,t) = Z$, with a
Lagrange multiplier $\lambda$.

The dissipative function $F$ in the Rayleighian is determined by
two relative velocities -- the movement of a chain segment with
respect to its environment
 \begin{equation}
\left. \left( \frac{{\rm d} \vec{r}}{{\rm d}t}- \frac{{\rm d}
\vec{R}}{{\rm d}t} \right) \right|_{\mbox{$s$:fixed}} =
\left({\vec{v}} - \left. \frac{\partial {\vec{R}}}{\partial n
}\right|_t v_n - \left.{
\partial {\vec{R}} \over \partial t }\right|_n \right)  \label{relv}
\end{equation}
and the slip motion of the chain segment equilibrium reference
point along the primitive path
 \begin{equation}
\left( \left.\frac{{\rm d}\vec{R}}{{\rm d}t
}\right|_{\mbox{$s$:fixed}} - \left. \frac{\partial
{\vec{R}}}{\partial t }\right|_n \right) = \left.{ \partial
{\vec{R}} \over \partial n }\right|_t v_n \ ,   \label{slip}
 \end{equation}
where the local velocity of polymer segment and the rate of the
primitive path evolution are given by, respectively, $$ {\vec{v}}
=\left. { \partial {\vec{r}} \over \partial t } \right|_s  \qquad
{\rm and} \qquad  v_n = \left. { \partial n \over \partial t }
\right|_s . $$
 Two different friction constants are then
introduced to describe the dissipation corresponding to these two
relative velocities, eqs.~(\ref{relv}) and (\ref{slip}), as they
appear in the dissipative functional $F$:
\begin{equation}
F = {1\over 2} \zeta \int_0^L {\rm d}s \left({\vec{v}} - \left.{
\partial {\vec{R}} \over \partial n }\right|_t v_n - \left.{
\partial {\vec{R}} \over \partial t }\right|_n \right)^2 +{1 \over 2}
\nu \int_0^L {\rm d}s \left( \left.{ \partial {\vec{R}} \over
\partial n }\right|_t v_n  \right)^2 , \label{F}
\end{equation}
The equations of motion are obtained from the Rayleighian ${\cal
R}$ through the variational equations
\begin{eqnarray}
- {{\delta F} \over {\delta \vec{v}}}- {{\delta U} \over {\delta
\vec{r}}} = 0 \qquad  {\rm and}  \qquad - {{\delta F} \over
{\delta {v_n}}}- {{\delta U} \over {\delta n}} = 0. \label{eq12}
\end{eqnarray}
The first of equations~(\ref{eq12}) gives, after evaluating the
functional derivatives,
\begin{equation}
\zeta\left({\partial \vec{r} \over \partial t} - {\partial \vec{R}
\over
\partial n}{\partial n \over \partial t}-{\partial \vec{R} \over
\partial t}\right)+ { 3 k_{\rm B} T \over b } \left[ -
{\partial^2 \vec{r} \over \partial s^2} + q_0^2 ( \vec{r} -
\vec{R} ) \right] = 0  . \label{eq1a}
\end{equation}
The second varuiational equation~(\ref{eq12}) gives
\begin{equation}
-\zeta {\partial \vec{R} \over \partial n} \cdot \left({\partial
\vec{r} \over \partial t} - {\partial \vec{R} \over \partial n}
{\partial n \over \partial t} - {\partial \vec{R} \over \partial
t} \right) + \nu \left( {\partial \vec{R} \over \partial n}
\right)^2 {\partial n \over
\partial t} + { 3 k_{\rm B} T \over b } q_0^2 ( \vec{r} - \vec{R} ) \cdot
\left( - {\partial \vec{R} \over \partial n} \right) = 0 ,
\nonumber
\end{equation}
which can be simplified to
\begin{equation}
\nu\left( {\partial \vec{R} \over \partial n} \right)^2 {\partial
n \over \partial t} +{ 3 k_{\rm B} T \over b } { \partial \vec{R}
\over
\partial n } \cdot \left[ -{\partial^2 \vec{r} \over \partial s^2}
\right]= 0 \label{eq2a}
\end{equation}
with the help of eq.~(\ref{eq1a}).

A natural dynamical variable in our problem is not the absolute
chain segment position $\vec{r} (s,t)$ but the position relative
to the primitive path, the fact we have already exploited in
eq.~(\ref{relv}) for the velocity. Accordingly, we introduce the
transverse excursion
\begin{equation}
\tilde{\vec{r}}(s,t)\equiv \vec{r} (s,t)- \vec{R} [n(s,t),t].
\label{rtilde}
\end{equation}
 The reason for using of the relative-position variable
$\tilde{\vec{r}}(s,t)$ is to separate the fast and slow
contributions to the dynamic equations. In particular, examine the
eq.~(\ref{eq2a}) for evolution of the primitive path: the equation
of motion for $n(s,t)$. If we neglect the effect of fast
transverse fluctuations around the primitive path,
$\tilde{\vec{r}}$, this equation transforms to:
\begin{equation}
\nu\left( {\partial \vec{R} \over \partial n} \right)^2 {\partial
n \over \partial t} + { 3 k_{\rm B} T \over b} {\partial \vec{R}
\over
\partial n} \cdot \left[-{{\rm d}^2 \vec{R} \over {\rm d} s^2}
\right] = 0 . \label{eq2c}
\end{equation}
After one identifies the full derivatives $${{\rm d} \vec{R} \over
{\rm d} s}= {\partial \vec{R} \over \partial n} {\partial n \over
\partial s} \ \ \ \ {\rm and} \ \ \ {{\rm d}^2 \vec{R} \over {\rm d}
s^2}= {\partial^2 \vec{R} \over \partial n^2} \left( {\partial n
\over
\partial s} \right)^2 + {\partial \vec{R} \over \partial n}
{\partial^2 n \over \partial s^2}  $$ and makes the according
substitution, the part of the dynamic equation~(\ref{eq2a})
describing the slow evolution of the primitive path takes the form
\begin{equation}
{\partial n \over \partial t} + { 3 k_{\rm B} T \over b \nu }
\left[ - {\partial \vec{R} \over \partial n} \cdot {\partial^2
\vec{R} \over \partial n^2} \left( {\partial \vec{R} \over
\partial n} \right)^{-2} \left( {\partial n \over \partial s}
\right)^2 - {\partial^2 n \over \partial s^2} \right] = 0 .
\label{eq2d}
\end{equation}
Note that, because $s$ is measured in units of length, the
dimensionality of $\nu$ (and $\zeta$) is such that the product $(b
\nu)$ is an ordinary friction constant measured, e.g., in
$\hbox{J\, s/m}^2$. In equilibrium, when no external deformation
is applied to the system, the primitive path $\vec{R}_0$ is given
by the static solution of the eq.~(\ref{eq2d})
\begin{equation}
-{\partial {\vec{R}}_0 \over \partial n} \cdot {\partial^2
{\vec{R}}_0 \over \partial n^2} \left( {\partial {\vec{R}}_0 \over
\partial n} \right)^{-2} \left( {\partial n_0 \over \partial s}
\right)^2 - {\partial^2 n_0 \over \partial s^2} =0.
\label{equilib1}
\end{equation}
Since the equilibrium step length of the primitive path $a$ is
assumed to be fixed, one finds
 \begin{equation}
\left| {\partial {\vec{R}}_0 \over \partial n} \right| = a \qquad
\rightarrow  \ \ {\partial {\vec{R}}_0 \over \partial n} \cdot
{\partial^2 {\vec{R}}_0 \over \partial n^2} = 0. \label{ortho}
\end{equation}
With this orthogonality relation, the equilibrium condition
(\ref{equilib1}) is simply
\begin{equation}
{\partial^2 n_0 \over \partial s^2} =0 \qquad \rightarrow \ \
n_0(s) = {Z \over L}s ,  \label{n0}
\end{equation}
a homogeneous distribution satisfying the boundary conditions
$n_0(0) = 0$ and $n_0(L) = Z$. Let us note here, although a more
extended discussion will follow later in the paper, that a limit
of completely uncrosslinked system requires a non-trivial
analysis. With respect to eq.~(\ref{n0}) one simply needs to
remember that $Z$ and $L$ are not independent parameters, in fact
a direct proportionality $Lb=Za^2$ holds. Therefore, even for melt
of uncrosslinked chains, or in the limit $L \rightarrow \infty$,
the constant ratio $L/Z$ is always reflecting the average degree
of entanglement, giving the arc distance between the constraints.

%--------------------------------------------------------------
\section{The linear response}
%--------------------------------------------------------------

We now consider the situation when an external deformation is
suddenly applied to the network, and calculate the linear dynamic
stress response. The principal feature of this analysis is the
computation of non-equilibrium evolution of the tube surrounding
the polymer chain. First, the primitive path trajectory, $\vec{R}
(n,t)$, determined by local restrictions imposed by other chains
in a dense system, is assumed to be deformed affinely with the
whole body:
\begin{equation}
\vec{R} (n,t)= \matr{E} (t) \cdot {\vec{R}}_0(n) \label{affine}
\end{equation}
with $\matr{E} (t)$ the deformation gradient tensor. For example,
in a fixed coordinate system, $\matr{E}$ is given by
\begin{equation}
\matr{E}_s= \left[\begin{array}{ccc}1& \lambda_{12} &0 \\ 0& 1 & 0
\\ 0 & 0 & 1
\end{array} \right]  \qquad {\rm and} \qquad \matr{E}_e=
\left[\begin{array}{ccc} \lambda_1 & 0 & 0 \\ 0 & \lambda_2 & 0 \\
0 & 0& \lambda_3 \end{array} \right] \label{strains}
\end{equation}
for simple shear deformation and elongation, respectively. Since
we shall be looking for the linear response, let us identify the
small strain $\matr{\epsilon} (t)$ through ${\mathsf E}_{\alpha
\beta} (t)= \delta_{\alpha \beta} + \epsilon_{\alpha \beta} (t)$.
For the simple shear, only one component of strain tensor is
present, $\epsilon_{\alpha \beta}=\vep \, u_\alpha g_\beta$, where
$\vec{u}$ is the displacement direction, $\vec{g}$ is that of a
gradient and the amplitude $\vep \ll 1$. For the extension, one
obtains $\epsilon_{\alpha \beta}=\vep_\alpha \delta_{\alpha
\beta}$. If the material is assumed fully incompressible, the
imposed principle extension $\vep \equiv \vep_3 \ll 1$ results in
the other two diagonal elements equal to each other by rotational
symmetry and $\vep_1= \vep_2= -\frac{1}{2}\vep$.

It is necessary to separate the small non-equilibrium correction
to the primitive path coordinate, $n_1$, defined by $ n(s,t) = n_0
(s) + n_1 (s,t)$. Substituting the affine deformation condition
(\ref{affine}) and using the fixed-step and orthogonality
relations (\ref{ortho}), the linearised dynamic equation
(\ref{eq2d}) takes the form of one-dimensional diffusion equation
for a scalar function $n_1(s,t)$:
\begin{equation}
{b \nu \over 3 k_{\rm B} T }{\partial n_1\over \partial t }-
{\partial^2 n_1 \over \partial s^2 }= \Psi(s,t), \label{eq2e}
\end{equation}
with the external force
\begin{equation}
\Psi(s,t)=  \left({Z \over a L} \right)^2  {\partial {\vec{R}}_0
\over \partial n } \cdot (\matr{\epsilon}{}^{\sf T} +
\matr{\epsilon}) \cdot {\partial^2 {\vec{R}}_0 \over
\partial n^2 } \equiv \left( \frac{L}{a^2Z} \right)\, \frac{\rm d}{{\rm d}s}
\left[ \frac{{\rm d}R_{0 \, \alpha}}{{\rm d}s} \epsilon_{\alpha
\beta} \frac{{\rm d}R_{0 \, \beta}}{{\rm d}s} \right] .
\label{fsdef}
\end{equation}

The diffusion equation of motion for the primitive path,
eq.~(\ref{eq2e}), is similar to what one frequently finds in the
statistical theory of polymers, cf. \cite{Doibook}. The diffusion
constant of the homogeneous equation is $D_n=3k_{\rm B}T/b \nu$.
In our case, since we shall be considering an oscillating imposed
strain and looking for a response at a given frequency $\omega$,
it is convenient to introduce the equilibrium Green's function
that satisfies
\begin{equation}
\left( \mi \, \omega {b \nu \over 3 k_{\rm B} T }-{{\rm d}^2\over
{\rm d} s^2} \right) G(s,s^\prime )= \delta (s-s^\prime )
\label{Geq}
\end{equation}
and is explicitly given by
\begin{equation}
G(s,s^\prime )= {1 \over 2\gamma } \left[\me^{- \gamma |s -
s^\prime |}- { 1 \over \sinh \gamma L } \left\{ \me^{- \gamma ( L
- s^\prime ) } \sinh \gamma s + \me^{- \gamma  s^\prime  } \sinh
\gamma (L - s) \right\} \right] , \label{G1}
\end{equation}
with the characteristic parameter, of the inverse-length
dimensionality,
\begin{equation}
\gamma= \sqrt{{\mi \, b \nu \, \omega \over 3 k_{\rm B} T } } \ .
\label{gamma}
\end{equation}
This Green's function satisfies the required chain reversal
symmetry and the boundary conditions $$G(s, s^\prime )= G(s^\prime
,s );  \qquad G( s,0 )= G(s,L )=0 . $$  The dynamical solution of
the eq.~(\ref{eq2e}) with the perturbation $\Psi(s,t)$ and the
boundary conditions $n_1 (0,t) = n_1 (L,t) = 0$ is then given by
the convolution
\begin{equation}
n_1 (s,t)= \int_0^L  {\rm d} s^\prime \, G(s, s^\prime )  \,
\Psi(s^\prime , t) \ . \label{n10sol1}
\end{equation}
The main dynamic equation~(\ref{eq2e}), with the tensile force
$\Psi(s,t)$ due to effective elongation of the primitive path
under deformation, lead to a non-trivial relaxation behaviour --
the key result of this paper.

Once the linear correction to the primitive path is obtained [$n_1
(s,t)$ generated by an imposed strain $\matr{\epsilon} (t)$], one
can compute the dynamic stress response. In a first approximation,
the macroscopic stress is calculated through the correlation
function along the chain (cf. \cite{Doibook}),
\begin{equation}
\sigma_{\alpha \beta}=c_{\rm x} {3 k_{\rm B} T \over b } \int_0^L
{\rm d}s \left\langle { \partial r_\alpha \over \partial s } {
\partial r_\beta \over \partial s } \right\rangle , \label{sdef}
\end{equation}
where $c_{\rm x}$ denotes the number density of polymer chains in
the network (a quantity directly related to the crosslink
density). In this equation, $\vec{r}(s)$ is the absolute position
of chain segment given by $\vec{r} (s,t) = \vec{R} [n(s,t),t] +
\tilde{\vec{r}}(s,t)$. By neglecting the small fast-fluctuating
variable $\tilde{\vec{r}}$ we obtain
\begin{equation}
\vec{r} (s)\approx {\vec{R}}_0[n_0(s)] + \matr{\epsilon}(t) \cdot
{\vec{R}}_0[n_0(s)] + {
\partial {\vec{R}}_0[n_0(s)] \over \partial n_0 } n_1(s,t) ,
\label{rsmall}
\end{equation}
up to the first order in $\matr{\epsilon}$. Accordingly, one
obtains three distinct contributions to the average stress tensor,
$$\sigma_{\alpha \beta}=\sigma_{\alpha \beta}^{(0)} +
\sigma_{\alpha \beta}^{(1)} + \sigma_{\alpha \beta}^{(2)},$$
corresponding to the undistorted primitive path ${\vec{R}}_0$ in
the eq.~(\ref{rsmall}), the equilibrium response to the imposed
strain $\matr{\epsilon}$ and the relaxation part determined by the
sliding motion of polymer segments along the tube $n_1(s,t)$,
respectively:
\begin{eqnarray}
\sigma_{\alpha \beta}^{(0)} &=& c_{\rm x} { 3 k_{\rm B} T \over b
} \int_0^L {\rm d}s \left\langle { {\rm d} R_{0\alpha} \over {\rm
d} s } { {\rm d} R_{0\beta} \over {\rm d} s } \right\rangle,
\label{Msigma0} \\
 \sigma_{\alpha \beta}^{(1)} &=& c_{\rm x} { 3
k_{\rm B} T \over b } \int_0^L {\rm d}s \left[ \left\langle { {\rm
d} R_{0\alpha} \over {\rm d} s } { {\rm d} R_{0\gamma} \over {\rm
d} s } \right\rangle \epsilon_{\beta \gamma}+ \left\langle { {\rm
d} R_{0\beta} \over {\rm d} s } { {\rm d} R_{0\gamma} \over {\rm
d} s } \right\rangle \epsilon_{\alpha \gamma} \right] ,
\label{Msigma1} \\
  \sigma_{\alpha \beta}^{(2)} &=& c_{\rm x} { 3
k_{\rm B} T \over b } \left(\frac{L}{Z} \right) \int_0^L {\rm d}s
\left\langle { {\rm d} R_{0\alpha} \over {\rm d} s } { {\rm d}
R_{0\beta} \over {\rm d} s } \frac{\partial n_1(s,t)}{\partial s}
\right\rangle . \label{Msigma2}
\end{eqnarray}
The first term is the small correction to the isotropic
hydrostatic pressure in the system, determined by thermodynamic
conditions (see, e.g. \cite{Doibook}); we shall not discuss it any
longer here. The second contribution, $\matr{\sigma}{}^{(1)}$, is
the equilibrium entropic response of a chain confined to the
primitive path random walk to an imposed strain; this is a
time-independent effect. All leading-order effects of relaxation
are concentrated in the third term, $\matr{\sigma}{}^{(2)}$, where
the movement of chain segments is calculated with the help of
equation~(\ref{n10sol1}).

%---------------------------------------------
\section{The complex modulus $G^*(\omega)$}
%---------------------------------------------

We now turn to particular cases of imposed small deformation and,
first, consider the oscillating simple shear of the sample, cf.
$\matr{E}_s$ of eq.~(\ref{strains}); the corresponding small
strain tensor is $\epsilon_{\alpha \beta}=\vep \, u_\alpha g_\beta
\, \me^{\mi \, \omega t}$. In order to evaluate averages in
equations (\ref{Msigma1}) and (\ref{Msigma2}) one needs to
recognize that, for the equilibrium primitive path
${\vec{R}}_0(n_0)$ one has, for instance,
\begin{equation}
\left\langle { {\rm d} R_{0\alpha} \over {\rm d} s } { {\rm d}
R_{0\beta} \over {\rm d} s } \right\rangle = \left( \frac{Z}{L}
\right)^2 \left\langle { {\rm d} R_{0\alpha} \over {\rm d} n } {
{\rm d} R_{0\beta} \over {\rm d} n } \right\rangle = \left(
\frac{Z}{L} \right)^2 \frac{1}{3} a^2 \delta_{\alpha \beta} ,
\label{pair}
 \end{equation}
the pair average of the primitive path tangent vector $\vec{t}
=({\rm d} {\vec{R}}_{0}/{\rm d} n)$, which is assumed constant for
all chain segments (labeled by $s$) belonging to the same step
$n$. In this way, the contour integral over the chain segments is
divided into a sum over steps of the primitive path: $$ \int_0^L
{\rm d}s = \sum_{n=1}^Z \int_{\frac{L}{Z}(n-1)}^{\frac{L}{Z}(n)}
{\rm d}s . $$ For the static stress $\matr{\sigma}{}^{(1)}$ this
division does not matter because the integrand in (\ref{Msigma1})
is not a function of $s$. One thus obtains
\begin{eqnarray}
 \sigma_{\alpha \beta}^{(1)} = c_{\rm x} { 3
k_{\rm B} T \over b } \ \frac{Z^2}{L}\frac{1}{3}a^2  \vep(t)
[u_\alpha g_\beta + u_\beta g_\alpha] \ = 2 c_{\rm x} Z \, k_{\rm
B} T \, \epsilon_{\alpha \beta}^{(S)} ,  \label{Msigma11}
\end{eqnarray}
where $\matr{\epsilon}{}^{(S)}$ is the symmetric part of the
imposed shear and the the relation $Lb=Za^2$ for the end to end
distances of a chain and a primitive path is used. This
equilibrium stress response is characterised by the effective
rubber modulus $\mu=2c_{\rm x} Z { k_{\rm B} T}$, proportional to
the total number of constraints for the average chain. Note that
because the chain of length $L$ is considered to have $(Z-1)$
entanglement points and one cross link, the product $c_{\rm x}Z$
can be written as $c_{\rm x}+c_{\rm e}$, involving the density of
entanglements $c_{\rm e}$.

Substituting the convolution relation (\ref{n10sol1}) with the
external force $\Psi(s,t)$ given by the eq.~(\ref{fsdef})  and
integrating by parts to switch the $s$-derivative, we obtain for
the relaxation part of stress:
\begin{eqnarray}
 \sigma_{\alpha \beta}^{(2)} = - c_{\rm x} { 3
k_{\rm B} T \over b } \left(\frac{L}{aZ} \right)^2 \int_0^L
\int_0^L {\rm d}s {\rm d}s' \frac{\partial^2 G}{\partial s \,
\partial s'} \left(\frac{Z}{L} \right)^4
\left\langle { {\rm d} R_{0\alpha} \over {\rm d} n } { {\rm d}
R_{0\beta} \over {\rm d} n } { {\rm d} R_{0\, a} \over {\rm d} n'
} { {\rm d} R_{0\, b} \over {\rm d} n' } \right\rangle
\epsilon_{ab} . \label{Msigma22}
\end{eqnarray}
For simple shear, with vectors  $\vec{u} \bot \vec{g}$, only the
four-term average gives a non-zero result, $\langle t_\alpha
t_\beta t_a t_b \rangle = \frac{1}{15}a^4 [\delta_{\alpha \beta}
\delta_{ab} + \delta_{\alpha a} \delta_{\beta b} + \delta_{\alpha
b} \delta_{\alpha b}]$ at $n=n'$. One thus obtains, after
replacing the integrals of full derivatives of the Green's
function $G(s,s')$,
\begin{eqnarray}
\sigma_{\alpha \beta}^{(2)} &=& - c_{\rm x} { 2 k_{\rm B} T \over
5 b } \left. \left. \left(\frac{a Z}{L} \right)^2 \epsilon_{\alpha
\beta}^{(S)} \sum_{n=1}^Z G(s,s')
\right|_{\frac{L}{Z}(n-1)}^{\frac{L}{Z}(n)}
\right|_{\frac{L}{Z}(n-1)}^{\frac{L}{Z}(n)} \label{Msigma222} \\
 &=&  - c_{\rm x} { 2 k_{\rm B} T \over
5 b } \left(\frac{a Z}{L} \right)^2 \frac{L}{\xi} f(\xi,Z) \
\epsilon_{\alpha \beta}^{(S)}(t) \nonumber ,
\\ {\rm where} && f(\xi,Z)= 1 - \me^{-\xi} - 2\me^{-Z\xi}{\rm csch}(Z\xi )\,
\sinh^2(\frac{\xi}{2})-\frac{1}{Z} \tanh (\frac{\xi}{2})
\label{fdef}
\\ {\rm and} && \xi=\frac{L \, \gamma}{Z} \equiv \frac{L}{Z}
\sqrt{\frac{i \, b\nu \, \omega}{3 k_{\rm B}T}} . \label{xdef}
\end{eqnarray}
For a simple fixed-frequency oscillating strain input, the linear
complex modulus $G^*(\omega)$ for shear deformation is defined by
the combination of the two relevant contributions to the stress
tensor:
\begin{equation}
G_s^*(\omega)= 2 c_{\rm x} Z \, k_{\rm B} T \left[1- {1 \over 5}{1
\over \xi } f(\xi,Z) \right]  . \label{G*}
\end{equation}
As should be expected for a classical rubber, one obtains exactly
the same expression for the linear modulus in response to the
imposed extension $\epsilon_{\alpha \beta}=\vep_\alpha
\delta_{\alpha \beta} \me^{\mi \, \omega t}$. That is, an explicit
calculation gives $G_s^*(\omega)=G_e^*(\omega)$, where the latter
is produced by the sum of two contributions to the normal stress,
\begin{eqnarray}
\sigma_{\alpha \beta}^{(1)} &=& 2 c_{\rm x} Z { k_{\rm B} T} \,
\vep_\alpha^0 \delta_{\alpha \beta} \, \Real{ \me^{\mi \omega t}}
, \label{sigma11f} \\ \sigma_{\alpha \beta}^{(2)} &=& -
\frac{2}{5} c_{\rm x} Z k_{\rm B} T \, \vep_\alpha^0
\delta_{\alpha \beta} \, \Real{ {1 \over \xi} f(\xi,Z) \, \me^{\mi
\omega t} }, \label{sigma22f}
\end{eqnarray}
where $c_{\rm x} Z=(c_{\rm x} +c_{\rm e})$ and the linear diagonal
extensions are related by $\vep_1 = \vep_2 = -\frac{1}{2}\vep_3$
if the fully incompressible case is considered.

\section{Discussion}
In eq.~(\ref{G*}), the complex modulus $G^*(\omega)$ is given as a
function of the complex non-dimensional parameter $\xi$, which is
defined in eq.~(\ref{xdef}). Since this parameter contains the
frequency $\omega$, the rest of it represents a time scale. This
characteristic time, which we denote $\tau_{\rm e}$, is determined
by the diffusion constant $D_n$ of the homogeneous equation of
motion (\ref{eq2e}) for $n_1$. For $D_n = 3k_{\rm B}T/ b \nu$ we
obtain
\begin{equation}
\tau_{\rm e}={1 \over 2 D_n} \left( { L \over Z } \right)^2 \ =
\frac{b \nu}{6 k_{\rm B}T} \left( \frac{a^2}{b} \right)^2 ,
\label{tauE}
\end{equation}
which is the time for a density fluctuation $n_1$ to diffuse a
distance $L/Z$, a step length of the primitive path or the arc
length of the polymer between the entanglements. This time scale
is also found in the theory of polymer melts, cf. \cite{Doibook},
where it separates the regimes of free Rouse dynamics and that
constrained by the tube. The relation to the Rouse time is
established when one recognises the product $(b \nu)$ as the
proper friction coefficient: $\tau_{\rm e}\simeq \tau_{\rm
R}/Z^2$.

The parameter $\xi$ can now be expressed as $\xi= (1+i) \sqrt{
\tau_{\rm e} \omega } $, which also means that $\tau_{\rm e}$ is
one of the characteristic time scales of the complex modulus
$G^*(\omega)$. Examining the eq.~(\ref{fdef}) one can identify the
second characteristic time appearing in the non-dimensional
exponent $Z \xi \sim \sqrt{\tau_{\rm R} \omega}$, with $\tau_{\rm
R} = L^2/2D_n$ the time for the slip-motion to diffuse all of the
chain contour length between the crosslinks.

The following limiting cases of non-dimensional function
$f(\xi,Z)$ can be easily seen from the definition (\ref{fdef}):
\begin{center}
\begin{tabular}{|c|c|c|c|c|}
\hline
 & $Z \rightarrow 1 \ (c_{\rm e}/c_{\rm x} \ll 1)$
& $Z\rightarrow \infty$ &$ \xi\rightarrow 0$ & $\xi \rightarrow
\infty$ \\
 \hline $ f(\xi,Z) \rightarrow$ \  & $\left(Z-1 \right)
\frac{ \xi + \sinh (\xi)}{1 + \cosh (\xi)}$ & $1-\me^{-\xi}$ & $
\xi(1-\frac{1}{Z})$ & $1-\frac{1}{Z}$ \\ \hline
\end{tabular}
\end{center}
which leads to a number of important limiting cases for the
storage and loss parts of the complex modulus $G^*(\omega)=G' +
\mi \, G''$~:
\begin{eqnarray}
({\rm arb.} \ Z) \qquad \qquad \qquad \qquad G'(\omega) \qquad
\qquad \qquad \qquad && \qquad G''(\omega) \label{Glimits} \\
 \tau_{\rm e}\omega \ll 1 :  \ \  \textstyle{\frac{8}{5}}c_{\rm x} k_{\rm B}T
\left[ \left(Z+\textstyle{\frac{1}{4}} \right) +
\textstyle{\frac{1}{90}}\, \left( Z-1 \right) Z^3 (\tau_{\rm
e}\omega )^2 \right] &,& \textstyle{\frac{2}{15}}\left( Z-1
\right)Z \, c_{\rm x} k_{\rm B}T \, \tau_{\rm e} \omega \nonumber
\\
 \tau_{\rm e}\omega \gg 1 : \ \ \qquad  2 Z c_{\rm x}
k_{\rm B}T -\textstyle{\frac{1}{5}}(Z-1) c_{\rm x} k_{\rm
B}T(\tau_{\rm e}\omega)^{-1/2} &,& \textstyle{\frac{1}{5}}(Z-1)
c_{\rm x} k_{\rm B}T \,(\tau_{\rm e}\omega)^{-1/2}   \nonumber
\end{eqnarray}

%%%%%%
\begin{figure} [h]
\centerline{\resizebox{0.84\textwidth}{!} {
\includegraphics{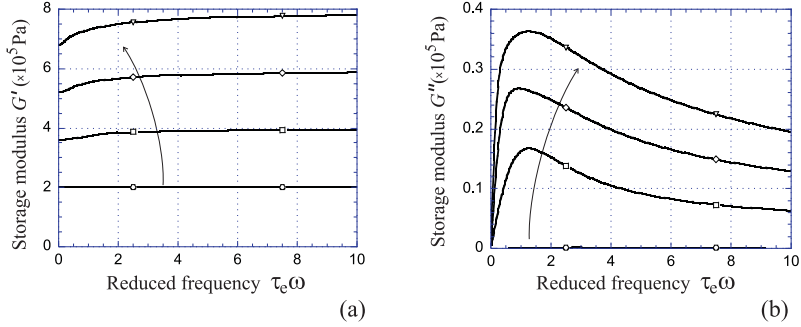} } }
\caption{(a) Plots of storage modulus $G'(\omega)$, in absolute
units, taking the value $c_{\rm x}k_{\rm B}T =10^5 \hbox{J/m}^3$
(a typical rubber-elastic energy scale), against $\tau_{\rm
e}\omega$ for $Z=$1,2,3,4 (increasing curves). The growth of the
zero-frequency value $\mu_0$ and a difference between it and the
plateau value at $\tau_{\rm e}\omega \gg 1$ are in evidence. \ (b)
Plots of loss modulus $G''(\omega)$, for the same set of
parameters and $Z$s, show an increasing effect of relaxation for
more entangled network strands. } \label{G12}
\end{figure}
\begin{figure} [h]
\centerline{\resizebox{0.84\textwidth}{!} {
\includegraphics{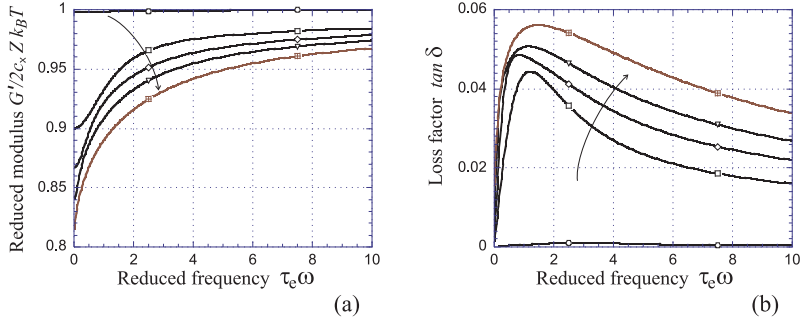} } }
\caption{(a) Plots of reduced storage modulus $G'/2 c_{\rm x} Z \,
k_{\rm B}T $ (non-dimensional) for $Z=$1,2,3,5 and 100 (decreasing
curves). \ (b) Plots of the loss factor $\tan \delta =G''/G'$, for
the same set of $Z$s (increasing curves). Note a slow decay of
$\tan \delta$ (and increase of $G'$) at $\tau_{\rm e}\omega \gg
1$, which is due to the square-root approach to saturation values
in eqs.~(\ref{Glimits}).} \label{Gred}
\end{figure}

%%%%%%%
The storage and loss moduli $G'$ and $G''$ are plotted against the
reduced frequency in Figs.~\ref{G12}(a,b), in absolute units:
taking a typical for rubbers value $c_{\rm x}k_{\rm B}T \sim 10^5
\hbox{Pa}$. An important relation between the storage modulus
value at zero frequency, the static modulus $\mu_0$, and that in
the high-frequency limit (the ``rubber plateau'' value at
$\tau_{\rm e} \omega \gg 1$) is characterised by a universal ratio
$$ \frac{ G'|_{\tau_{\rm e}\omega \gg 1}}{G'|_{\omega \rightarrow
0}} = \frac{5Z}{4Z+1} \ = \frac{1+(c_{\rm e}/c_{\rm
x})}{1+\frac{4}{5}(c_{\rm e}/c_{\rm x})}. $$
 Figure~\ref{Gred}(a) makes this point more explicit by plotting
the non-dimensional reduced storage modulus $G'/2c_{\rm x}Z \,
k_{\rm B}T$, cf. eq.~(\ref{G*}). At $Z \rightarrow 1$ there is no
variation of $G'$ with frequency. There is also no dissipation in
this limit of high dilution, as shown by the plots of loss factor
$\tan \delta$ in Fig.~\ref{Gred}(b). When entanglements dominate
the rubber-elastic response, at $Z \gg 1$ (perhaps a more
realistic case in typical rubber) one finds a 20\% increase in the
storage modulus for deformations on a time scale shorter than
$\tau_{\rm e}$. There is a substantial dissipation effect in this
limit of high entanglement, with the high-frequency tail of $\tan
\delta (\omega)$ decaying rather slowly, according to the
square-root asymptotic dependence, cf. eqs.~(\ref{Glimits}).
%%%%%%%%%%%
%%%%%%%%%%%%%%%%%%
Note that, the characteristic time of network strand relaxation is
rather short: in the reptation theory of melt dynamics one obtains
the hierarchy of time scales $\tau_{\rm e} \ll \tau_{\rm R} \ll
\tau_{\rm d}$ (the disengagement time, not present in an ideal
crosslinked network) with $\tau_{\rm e} \simeq \tau_{\rm R}/Z^2$
and $\tau_{\rm d} \simeq 3Z \tau_{\rm R}$. Taking just a few
recent cases \cite{callaghan}, out a large number of rheological
studies of polymer melts, one obtains, very crudely, for
polystyrene (PS) $\tau_{\rm e} \sim 10^{-7} \hbox{s}$ and for
poly(dimethyl) siloxane (PDMS) $\tau_{\rm e} \sim 10^{-8}
\hbox{s}$.
%%%%%%%%
\begin{figure} %[b]
\centerline{\resizebox{0.42\textwidth}{!} {
\includegraphics{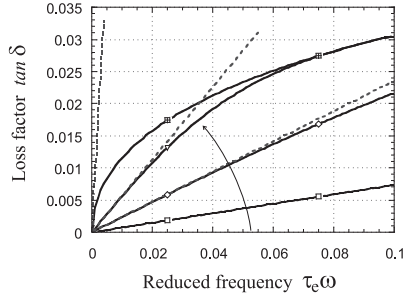} } }
\caption{Plots of the loss factor $\tan \delta $, for $\tau_{\rm
e} \omega \ll 1$ and the increasing values of $Z=2,4,8$ and $100$.
At moderate $Z$ the linear regime is observed (the calculated
slope is plotted as a dashed line for each $Z$). At large $Z \gg
(\tau_{\rm e} \omega )^{-1/4}$ one finds an apparent square-root
scaling in a wide range of low frequencies.} \label{tanD}
\end{figure}
%%%%%%%%%%%%%%%%%%

As a result, in some experiments, it may be difficult to access
the high-frequency end of the predicted relaxation behaviour. It
is, therefore, interesting to examine the asymptotic low-frequency
behaviour of the loss factor for the cases of $Z \geq 1$ and $Z
\gg 1$, when the number of entanglements per network strand
becomes large. Equations (\ref{Glimits}) indicate that at
$\tau_{\rm e}\omega \rightarrow 0$ the loss factor is linear:
$\tan \delta \approx (Z-1)Z/(12Z+3) (\tau_{\rm e}\omega )$.
However, as $Z$ increases so that $Z \gg (\tau_{\rm e} \omega
)^{-1/4}$, a different regime becomes apparent -- see
Fig.~\ref{tanD}. In fact, this intermediate regime $1 \gg
\tau_{\rm e} \omega \gg 1/Z^4$ is best fit by an effective
power-law $\tan \delta \approx 0.09 (\tau_{\rm e} \omega)^{0.45}$,
nearly another square-root.

The present calculation produces a static limit ($\omega
\rightarrow 0$) which is exactly equal to the result of Doi and
Edwards \cite{Doibook}, derived by assuming that the deformation
of the primitive path is affine and that the tensile force is
constant along the chain. Such a correspondence occurs because (i)
the polymer density along the tube is uniform for $\omega =0$ and
(ii) in the presence of crosslinks at the both ends of the polymer
there can neither be the contour length relaxation nor the
disengagement relaxation.

Ball, Doi, Edwards and Warner (BDEW) \cite{sliplink} derived the
formula for the elastic free energy  within the sliplink model for
entangled polymer networks, which produced the following
expression of the static modulus:
\begin{equation}
G'= 2k_{\rm B}T \left[ N_c + N_s { 1 \over \left( 1 + \eta
\right)^2 } \right], \label{BDEW}
\end{equation}
where, in the notation of \cite{sliplink}, $N_c$ and $N_s$ denote
the numbers of crosslinks and slip links, respectively. An
additional parameter $\eta$ characterises the amount of chain
which can slip through a sliplink and, according to BDEW, is given
by
\begin{equation}
\eta \sim {4 \over 3} \left( \frac{(N_c+N_s) \Delta}{L} \right)^2.
\end{equation}
Here, $ L/ \left( N_s+N_c \right)$ is the arc distance between two
sliplinks, which is analogous to $L/Z$, the equilibrium step
length of the primitive path, in the present paper, and $\Delta$
is the allowed slip arc length. By identifying that $c_{\rm x} =
N_c$ and $c_{\rm e} = N_s$ for unit volume, we can see that the
limits $G'(\tau_{\rm e}\omega \gg 1)$ and $G'(\tau_{\rm e}\omega
\ll 1)$ of the present calculation correspond to the following
values of $\Delta$ in the BDEW result (\ref{BDEW}): At $\omega
\rightarrow \infty$ one obtains $\Delta = 0$ (no slipping past the
constraint) and at $\omega \rightarrow 0$ the slipping occurs to
the amount $\Delta \sim 0.3 (L/Z)$.

One should note that a limiting case of zero crosslinking density,
a corresponding entangled polymer melt, is not trivially achieved
by setting $c_{\rm x} \rightarrow 0$ in the consistent dynamical
theory. At sufficiently short times, or frequency above $\tau_{\rm
d} \omega \sim 1$, the dynamic mechanical response is controlled
by the chain constraints which are assumed to have the fixed
density $c_{\rm e} \sim Z/Ld^2$ (for $c_{\rm x} \sim 1/v_{\rm
chain} = 1/Ld^2$ with $d^2$ the cross-section). This, and a
higher-frequency regime when $\tau_{\rm e} \omega \geq 1$, is
where the present model of sliding motion along the tube finds its
application. The regime of chain disentanglement in a
corresponding melt could be achieved, in a simple way, by setting
phenomenologically an effectively decreasing entanglement density
$c_{\rm e}^{\rm eff} \sim \exp \left( -t/\tau_{\rm d} \right)$,
vanishing completely at $t \gg \tau_{\rm d}$ due to reptation, see
\cite{Doibook}.

To conclude, we have developed a first-approximation description
of polymer dynamics in an ideal permanently crosslinked network
under small deformations. This results in a general expression for
linear complex modulus $G^*(\omega)$, which shows a characteristic
single-relaxation time behaviour in its frequency variation. The
low-frequency limits of storage and loss moduli have the classical
dependence on frequency, square and linear, respectively. We
obtain a rather slow, square-root variation of $G'$ and $G''$ in
the limit of high frequency, at $\tau_{\rm e} \omega \gg 1$. The
relaxation time scale $\tau_{\rm e}$, resulting from the
description in terms of one-dimensional diffusion of chain
segments along the distorted primitive path, is a quantity closely
related to the characteristic entanglement (tube formation) time
in a corresponding melt, although the topological sense (and,
hence, the density of) entanglements changes when the chains are
permanently crosslinked into a network.

One of the principal achievements of this work is its consistent
use of Rayleighian dissipative function approach. This method is
naturally adopted for the treatment of fluctuations, relaxation
and constraints, and could be useful for many other problems of
polymer dynamics.

%-----------------------------------------------------------

%-----------------------------------------------------------------------
\newpage
\begin{center} FIGURE CAPTIONS \end{center}
FIGURE 1. \ \ \ (a) The scheme of tube model and the relevant
variables: the polymer conformation $\vec{r}(s,t)$ and the tube
primitive path $\vec{R}(n,t)$, where the function $n(s,t)$
specifies the position along the fixed-length path and $s$
specifies the position along the chain. In a network, the chain is
permanently crosslinked at both ends of the tube. \ \ (b) The
``expected'' behaviour of the dynamic storage modulus
$G'(\omega)$: For an uncrosslinked melt the static modulus is
zero, but reaches a rubber plateau at a finite frequency due to
entanglements. On increasing concentration of permanent crosslinks
the static modulus $G'|_{\omega=0}=\mu_0$ increases; the plateau
modulus has contributions from both crosslinks and entanglements.
At a much higher frequency all curves should converge to the glass
plateau, which is not the subject of this paper. \\

FIGURE 2. \ \ \ (a) Plots of storage modulus $G'(\omega)$, in
absolute units, taking the value $c_{\rm x}k_{\rm B}T =10^5
\hbox{J/m}^3$ (a typical rubber-elastic energy scale), against
$\tau_{\rm e}\omega$ for $Z=$1,2,3,4 (increasing curves). The
growth of the zero-frequency value $\mu_0$ and a difference
between it and the plateau value at $\tau_{\rm e}\omega \gg 1$ are
in evidence. \ (b) Plots of loss modulus $G''(\omega)$, for the
same set of parameters and $Z$s, show an increasing effect of
relaxation for more entangled network strands. \\

FIGURE 3. \ \ \ (a) Plots of reduced storage modulus $G'/2 c_{\rm
x} Z \, k_{\rm B}T $ (non-dimensional) for $Z=$1,2,3,5 and 100
(decreasing curves). \ (b) Plots of the loss factor $\tan \delta
=G''/G'$, for the same set of $Z$s (increasing curves). Note a
slow decay of $\tan \delta$ (and increase of $G'$) at $\tau_{\rm
e}\omega \gg 1$, which is due to the square-root approach to
saturation values in eqs.~(\ref{Glimits}).\\

FIGURE 4. \ \ \ Plots of the loss factor $\tan \delta $, for
$\tau_{\rm e} \omega \ll 1$ and the increasing values of $Z=2,4,8$
and $100$. At moderate $Z$ the linear regime is observed (the
calculated slope is plotted as a dashed line for each $Z$). At
large $Z \gg (\tau_{\rm e} \omega )^{-1/4}$ one finds an apparent
square-root scaling in a wide range of low frequencies.
\end{document}